\documentclass[prl,nofootinbib,preprint,superscriptaddress]{revtex4}

\usepackage{graphicx}
\usepackage{amsmath}
\usepackage{amsfonts}
\usepackage{amssymb}
\usepackage{dsfont}
\usepackage{dcolumn}
\usepackage{bm}
\usepackage{amsmath,amssymb}
\usepackage{graphicx}
\usepackage[pdftex]{hyperref}
\usepackage{pstricks}
\usepackage{color}

\newcommand{\beqn}{\begin{eqnarray}}
\newcommand{\eeqn}{\end{eqnarray}}
\newcommand{\be}{\begin{equation}}
\newcommand{\ee}{\end{equation}}

\newcommand{\mathsym}[1]{{}}

\begin{document}

\title{$f(T)$ theories from holographic dark energy models}
\author{Peng Huang}

\affiliation{Institute of Theoretical Physics, Beijing University of Technology\\ Beijing 100124, China}
\author{Yong-chang Huang}

\affiliation{Institute of Theoretical Physics, Beijing University of Technology\\ Beijing 100124, China}
\affiliation{Kavli Institute for Theoretical Physics, Chinese Academy of Sciences\\ Beijing 100080, China}
\affiliation{CCAST (World Lab.), P.O. Box 8730\\ Beijing 100080, China}




\begin{abstract}
We reconstruct $f(T)$ theories from three different holographic dark energy models in different time durations. For the HDE model, the dark energy dominated era with new setting up is chosen for reconstruction, and the radiation dominated era is chosen when the involved model changes into NADE. For the RDE model, radiation, matter and dark energy dominated time durations are all investigated. We also investigate the limitation which prevents an arbitrary choice of the time duration for reconstruction in HDE and NADE, and find that an improved boundary condition is needed for a more precise reconstruction of $f(T)$ theory.

\noindent \it{PACS: 04.50.Kd, 95.36.+x, 98.80.-k}\\
\noindent \it{Keywords: Holography dark energy models; $f(T)$ theory; reconstruction}

\end{abstract}

\maketitle

\section{1. Introduction}
\label{sec:intro}

Cosmological observations from Type Ia Supernovae \cite{SN}\cite{riess}, cosmic microwave
background (CMB) radiation \cite{Spergel}\cite{Komatsu}, large scale structure (LSS) \cite{T}\cite{j}, baryon acoustic oscillations (BAO)\cite{D}, and weak lensing \cite{B} denote that the universe is in a accelerating phase. Since ordinary matters always attract each other, one can only expect a decelerated expansion of the universe, thus, the acceleration denotes that there may be an exotic energy component with repulsive gravity, or, the theory of gravity itself should be modified in the cosmological scale. Under this consideration, approaches developed in order to explain how this accelerating occurs and what end the universe evolves towards can be divided into two different representative categories.

The first is introducing a new energy form, named dark energy, with an appropriate index of equation of state (EOS) $\omega$ to trigger the acceleration. Well-studied models of this type are models with rolling fields, such as Quintessence, Phantom, Quintom, K-Essence , and Chaplygin gas, to name a few. Recently,  some attempts have been made to probe the nature of dark energy according to the holographic principle \cite{susskind}\cite{t}. These resulting holographic dark energy models are also belong to the first type since they also attribute the acceleration to a new energy form. However, comparing with the character of appearance of exotic matter fields in models mentioned above, holographic dark energy models, which define dark energy by geometrical objects such as event horizon $R_h$ (HDE model \cite{li}\cite{li2}), cosmological conformal time $\eta$ (NADE model \cite{cai1}\cite{cai2}) and the Ricci scalar $R$ (RDE model \cite{gao}), are totally different because of the absence of exotic matter fields.

The second category of theories, on the other hand, needn't to introduce exotic energy form but modify the theory of gravity itself, such as $f(R)$, MOND, and DGP. $f(R)$ is one of the simplest modification of General Relativity (GR) in which the Lagrangian density is an arbitrary function of $R$. However, one can also consider ``teleparallelism'' with the Weitzenbock connection, which has torsion scalar $T$, not the curvature $R$, to define the theory of gravity \cite{6}-\cite{9}. The result is a theory equivalent to GR known as the Teleparallel Equivalent of General Relativity (TEGR). In order to explain both the early inflation and the late time acceleration of the universe, the teleparallel Lagrangian density also has been extended to a function of $T$ as $f(T)$ \cite{10}\cite{11}, similar to the spirit of extending from $R$ to $f(R)$. For recent discussions of the theoretical aspects of $f(T)$, see \cite{12}-\cite{16}.

Both the holographic dark energy models and the modified theory of GR relate the geometry of spacetime and its evolution tightly. Their common relations to geometry of the spacetime have inspired recent efforts of reconstructing $f(R)$ and $f(T)$ gravities, which can accommodate a accelerated universe, from holographic dark energy models \cite{17}-\cite{20}. In these reconstructions, one can see that the attention is on dark energy dominated epoch, and a phantom-like behavior of the Hubble parameter $H$ with future singularity is always assumed, which is naturally based on the constraints from cosmological observations (see \cite{rip2}-\cite{rip4} and reference therein).

From a theoretical point of view, the holographic dark energy model which involves into a de Sitter universe is of particular interest since it is required by the second law of thermodynamics \cite{li}, thus it is valuable to do the reconstruction from a universe with no future singularity but evolves into a de Sitter phase. Furthermore, because of the importance of radiation and matter dominated epoches in the cosmological study on both theoretical and phenomenological aspects, it's also important to see what constraints can holographic dark energy models bring to the corresponding $f(T)$ theory during these two time durations. These considerations motivate us to investigate $f(T)$ gravities from various holographic dark energy models with new setting up of the Hubble parameter and in new time durations.

The paper is organized as follows: in Sec. II, we briefly review $f(T)$ theory; in Sec. III, we reconstruct $f(T)$ theories with new assumptions of the form of the Hubble parameter according to HDE, NADE and RDE respectively, and new results will be given out. Finally, summary and discussion are given in Sec.IV.

\section{2. Brief review of $f(T)$ theory}

In the framework of $f(T)$ theory, Lagrangian density is extended from the torsion scalar $T$ to a general function $f(T)$ \footnote{See \cite{tegr} for a detail exposition of TEGR.}, similar to the story happened in $f(R)$ theories. And the action of modified
teleparallel gravity is given by \cite{10}\cite{11}\cite{bengochea2}
\begin{equation}
I =\frac{1}{2k^2}\int {\rm d}^4
x~e~\Big[f(T)+L_m\Big],\label{action}
\end{equation}
where $k^2=M_P^{-2}=8\pi G$, $e={\rm det}(e^i_{\mu})=\sqrt{-g}$, and $L_m$ is the Lagrangian density of matter.
The field equations of $f(T)$ modified
teleparallel gravity can be given by taking the variation of the action (\ref{action}) with respect to
the vierbein, one can obtain the field equations in $f(T)$ modified
teleparallel gravity as \cite{10}
\begin{equation}
S_i^{~\mu\nu}\partial_{\mu}
(T)f_{TT}(T)+\Big[e^{-1}\partial_{\mu}(eS_i^{~\mu\nu})-e_i^{\lambda}T^{\rho}_{~\mu\lambda}S_{\rho}^{~\nu\mu}\Big]f_T(T)
+\frac{1}{4}e_i^{\nu}f(T)=\frac{k^2}{2}e_i^{~\rho}T_{\rho}^{~\nu},\label{fTeqs}
\end{equation}
where subscript $T$ denotes a derivative with respect to $T$,
$S_i^{~\mu\nu}=e_i^{~\rho}S_{\rho}^{~\mu\nu}$ and $T_{\mu\nu}$ is
the matter energy-momentum tensor.

For a spatially-flat FRW metric for the universe with $T_\nu^\mu={\rm diag}(-\rho,p,p,p)$,
\begin{equation}
T=-6H^2,\label{T}
\end{equation}
where $H=\dot{a}/a$ is the Hubble parameter, also, \ref{fTeqs} reduces into \cite{10}

\begin{equation}
12H^2f_T(T)+f(T)=2k^2\rho,\label{fT1}
\end{equation}
\begin{equation}
48H^2\dot{H}f_{TT}(T)-(12H^2+4\dot{H})f_T(T)-f(T)=2k^2p.\label{fT2}
\end{equation}
Here $\rho$ and $p$ are the total energy density and pressure of the
matter inside the universe, respectively, and satisfy the
conservation equation
\begin{equation}
\dot{\rho}+3H(\rho+p)=0.\label{conteq}
\end{equation}
For the convenience of calculation, one can assume the units $8\pi G=1$ and decompose $f(T)$ into $f(T)=g(T)+T$ as that in \cite{18}, in this setting up, \ref{fT1} and \ref{fT2} turn into
\begin{equation}
3H^2=\rho-\frac{1}{2}g-6H^2g_T,\label{ft3}
\end{equation}
\begin{equation}
-3H^2-2\dot{H}=p+\frac{1}{2}g+2(3H^2)+\dot{H})g_T-24\dot{H}H^2g_{TT},\label{ft4}
\end{equation}
Comparing \ref{ft3} and \ref{ft4} with ordinary Friedmann equations that
\begin{equation}
3H^2=\rho+\rho_D,
\end{equation}
\begin{equation}
-3H^2-2\dot H=p+p_D,
\end{equation}
one gets relations between $\rho_D$, $p_D$ and $g(T)$ \cite{18}
\begin{equation}
\rho_D=-\frac{1}{2}g-6H^2g_T,\label{gt1}
\end{equation}
\begin{equation}
p_D=\frac{1}{2}g+2(3H^2+\dot H)g_T-24\dot H H^2g_{TT},\label{gt2}
\end{equation}
which can be combined into
\begin{equation}
\rho_D+p_D=\rho_D(1+\omega_D)=2\dot H g_T-24\dot H H^2g_{TT}.\label{gt3}
\end{equation}
It's apparent that, in order to solve this differential equation of $g(T)$ with respect to $T$, $\rho_D$, $\omega_D$ and $\dot H$ should be expressed as function of $T$, respectively.

\section{3. The reconstruction}

The key argument of holographic dark energy models is, due to the limit made by the formation of a black hole, an ultraviolet (UV) cut-off has to be related to an infrared (IR) cut-off, thus, if density of dark energy $\rho _D $ is the quantum zero-point energy density caused by the UV cut-off, the total energy in a region of size $L$ should not exceed the mass of a black hole of the same size, that is $L^3\rho _D \le LM_p^2 $, then one has $\rho _D =3C^2M_p^2 L^{-2}$,
here, $C$ is a numerical constant introduced for convenience and $M_p $ is Planck mass \cite{Cohen}. The next step of model construction is to find out a suitable IR cut-off that can accelerate the universe. There are many different choices for doing this, three of them (HDE, NADE and RDE) are of special interest in this paper.

The logic in reconstruction of $f(T)$ gravity from holographic dark energy is considering that the holographic dark energy is described by the modification of the gravity with respect to TEGR effectively, i.e., the dark energy density in \ref{gt1}, \ref{gt2} and \ref{gt3} should be expressed through holographic dark energy models, then, solve these equations and get the concrete form of function $f(T)$ with respect to the torsion scalar $T$. Thus, the progress of reconstruction is totally model dependent. As in what following, we will continue our reconstruction according to three different holographic dark energy models.

\subsection{3.1 $f(T)$ from HDE}

Event horizon $R_h=a\int_t^\infty{dt\over a} = a\int_a^\infty {da \over Ha^2}$ is used for IR cut-off in this model, the density of dark energy is then becomes (from now on, we set $8\pi G=1$)
\begin{equation}
\rho_{D} = 3c^2 R_h^{-2},\label{rho}
\end{equation}
The state index of this model is
\begin{equation}
\omega_D = -{1\over 3}-{2\over 3HR_h} = -{1\over 3}-{2\sqrt{\Omega_D}\over 3c},\label{omega}
\end{equation}
and the evolution equation of $\Omega_D$ is
\begin{equation}
{\Omega'_D\over\Omega_D}=(1-\Omega_D)\left(1+{2\sqrt{\Omega_D}\over c}\right).\label{OOmega}
\end{equation}
\ref{omega} and \ref{OOmega} are important results which tell the evolution and destiny of the universe. If one takes $c$ as a live parameter of the model, constraints from cosmological observations imply a phantom future of the universe, the reconstruction of $f(T)$ theory in this situation is studied in detail already, as having been pointed out in the introduction. We here want to investigate this problem purely from theoretical point of view, thus, as discussed detail in \cite{li}, $c=1$ is favored. Then, \ref{omega} and \ref{OOmega} together denote that $\omega\rightarrow -1$ and $\Omega\rightarrow 1$ when $a\rightarrow\infty$, that is, the dark energy will dominate the universe and evolve into the de Sitter phase, in this situation, the form of Hubble parameter must be taken as
\begin{equation}
H=\Lambda,\label{h1}
\end{equation}
with $\Lambda$ a constant. In this setting up, $\dot H =0$, and $\rho_D =-\frac{c^2T}{2}$, \ref{gt1} and \ref{gt2} are in fact equivalent to each other and turn into
\begin{equation}
Tg_T-\frac{1}{2} g+\frac{c^2}{2}T=0,
\end{equation}
whose solution is
\begin{equation}
g(T)=-c^2T+C_1T^{\frac{1}{2}},\label{solutionHDE}
\end{equation}
which $C_1$ is an integration constant. Thus, the result of reconstruction of $f(T)$ theory from HDE in a dark energy dominant era is
\begin{equation}
f(T)=(1-c^2)T+C_1T^{\frac{1}{2}}.\label{solution11}
\end{equation}
The integration constant $C_1$ can be determined by using the boundary condition $f(T)_{t=0}=T_0$ \cite{18}, similar to that in $f(R)$ theory \cite{boundarycondition} which is chosen on the basis of physically consideration. The final result is
\begin{equation}
f(T)=(1-c^2)T+c^2\sqrt {T_0T}.
\end{equation}
The situation we consider here is dark energy dominated era only, considering the integral character of the definition of the event horizon in HDE, it is apparat that a precise reconstruction $f(T)$ theory from HDE with respect to the full evolution of the universe, i.e., not only in dark energy dominated era but also in radiation and matter dominated era, needs a complete information about the evolution of the scale factor $a(t)$, which is not available at moment.

\subsection{3.2 $f(T)$ from NADE}

The conformal cosmological time $\eta = \int {dt \over a} = \int {da \over a^2 H}$ is used for the IR cut-off in this model, then, the density of dark energy is
\begin{equation}
\rho_D = {3n^2 \over \eta ^2}.\label{cai1}
\end{equation}
The state index of this model is
\begin{equation}
\omega_D = -1+{2\over 3n}{\sqrt{\Omega_D}\over a}.\label{cai2}
\end{equation}
The IR-cut off in NADE, the cosmological conformal time, is again a global defined quantity whose precise form needs the complete knowledge of the scale factor, so, because the lack of information, it is difficult to give a precise result of the integration $\eta = \int {dt \over a} = \int {da \over a^2 H}$ from the radiation dominant era to time when dark energy prevails. If one wants to get some valuable results, some approximation is needed. Under this consideration, we choose to investigate the time duration when radiation is prevail to other energy components in the universe, then, $a(t)=bt^{\frac{1}{2}}$ ($b$ is a positive constant) and $H=\frac{1}{2t}$, combining \ref{T}, it's easy to find
\begin{equation}
\dot H =\frac{T}{3},\label{adeh}
\end{equation}
\begin{equation}
\rho_D=\frac{n^2b^2}{4}(-6T)^{\frac{1}{2}}.\label{aderho}
\end{equation}
Inserting \ref{adeh}, \ref{aderho} and $\omega_D=-\frac{1}{3}$ into \ref{gt3}, one gets
\begin{equation}
T^2g_{TT}+\frac{T}{2}g_T-\frac{n^2b^2}{8}\sqrt {-6T}=0,\label{aderesult}
\end{equation}
with solution
\begin{equation}
g(T)=\frac{n^2b^2}{4}\ln(T)\sqrt {-6T}-\frac{n^2b^2}{2}\sqrt {-6T}+2C_1T^{\frac{1}{2}}+C_2,\label{adegt}
\end{equation}
which tells
\begin{equation}
f(T)=T+\frac{n^2b^2}{4}\ln(T)\sqrt {-6T}-\frac{n^2b^2}{2}\sqrt {-6T}+2C_1T^{\frac{1}{2}}+C_2,\label{adeft}
\end{equation}
Contrary to the situation in HDE, now one can use the boundary condition that $(\frac{df(T)}{dt})_{t=0}=(\frac{dT}{dt})_{t=0}$ and $f(T)_{t=0}=T_0$ \cite{18}, then, one gets
\begin{equation}
C_1=-\frac{\sqrt{-6}n^2b^2}{8}\ln (T_0),
\end{equation}
\begin{equation}
C_2=\frac{n^2b^2}{2}\sqrt{-6T_0}.
\end{equation}
$f(T)$'s form then turns into
\begin{equation}
f(T)=T+\frac{n^2b^2}{4} \ln(\frac{T}{T_0})\sqrt {-6T}-\frac{n^2b^2}{2}\sqrt {-6T}+\frac{n^2b^2}{2}\sqrt{-6T_0}.\label{adeftfinal}
\end{equation}

An interesting thing is, it has been proved that $\Omega_D \propto n^2b^2$ \cite{cai2} during radiation-dominated epoch, thus, in the $\Omega_D\rightarrow 0$ limit, \ref{adeftfinal} indeed turns into $f(T)=T$, which incorporates the boundary conditions well and shows the consistency of reconstruction $f(T)$ theory from NADE in radiation-dominated era.

\subsection{3.3 $f(T)$ from RDE}

The energy density of RDE in a flat universe reads
\begin{equation}
\rho_D=3\alpha(2H^2+\dot H).\label{rdeenergydensity}
\end{equation}
Its dynamical equation of state index $\omega_D$ is
\begin{equation}
{\omega_D}=\frac{1}{3}(1-\frac{2}{\alpha})\frac{1}{1+\frac{\alpha \rho_{m0}}{3(2-\alpha)C}e^{(1-\frac{2}{\alpha})\ln a}},
\end{equation}
which $C$ is an integration constant \cite{kkk}. For $\alpha \ll 2$ and $\ln a\gg 1$, one has an approximately constant equation of state that
\begin{equation}
\omega_D \approx \frac{1}{3}(1-\frac{2}{\alpha}).\label{omegaga}
\end{equation}

One can see that the value of $\alpha$ is crucial in determining the evolutionary
behavior of RDE. If $\alpha=\frac{1}{2}$, the dark energy evolves into the cosmological constant with the
expansion of the universe, such that ultimately the universe
will enter the de Sitter phase in the far future. When $\alpha<\frac{1}{2}$,
the RDE will exhibit a quintom-like evolution behavior which leads to a cosmic big rip \cite{rip}. According to the observational constraints from the joint
analysis of data of SN, BAO, and WMAP5 in \cite{rip2}, the best-fit result for $\alpha$ is $\alpha=0.359^{+0.024}_{-0.025}$, which shows that RDE will more likely behave as a phantom energy, it is expected that the big rip will occur in a finite time. The appearance of future singularity is the key difference of RDE from HDE and NADE.

The reconstruction of $f(T)$ theories from HDE and NADE is restricted in special time duration in the previous two sections. In fact, when one considers doing the reconstruction in an alternative time duration, such as matter dominated time in HDE or dark energy dominated time in NADE, difficulties in concrete calculation will arise due to the global character of the definition of dark energy in these two holographic dark energy, such as the relation between $\rho_D$ and scalar torsion $T$ is hard to be established which makes it difficult to solve \ref{gt1}, \ref{gt2} and \ref{gt3}. Contrary to HDE and NADE, the definition of dark energy density in RDE doesn't depend on the evolution information of the scale factor, which gives convenience to search for a more precise reconstruction of $f(T)$ theory from RDE model.

\subsubsection{3.3.1 Radiation dominated era}

In the radiation dominated era, $a(t)=bt^{\frac{1}{2}}$ ($b$ is a positive constant) and $H=\frac{1}{2t}$, combining \ref{T} and \ref{rdeenergydensity}, it's easy to find
\begin{equation}
\dot H =\frac{T}{3},
\end{equation}
\begin{equation}
\rho_D=3\alpha(2H^2+\dot H)=0.\label{rderho}
\end{equation}
Thus, \ref{gt3} turns into
\begin{equation}
Tg_{TT}+\frac{1}{2}g_T=0,
\end{equation}
whose solution is
\begin{equation}
g(T)=C_1\sqrt T +C_2,
\end{equation}
then, one gets
\begin{equation}
f(T)=T+C_1\sqrt T +C_2.\label{rder}
\end{equation}
By using the boundary conditions that $(\frac{df(T)}{dt})_{t=0}=(\frac{dT}{dt})_{t=0}$ and $f(T)_{t=0}=T_0$ \cite{18}, one gets the final form of $f(T)$ in radiation dominated epoch that
\begin{equation}
f(T)=T,
\end{equation}
which is consistent with the fact that the RDE dark energy density is zero (see \ref{rderho}) in this time duration, thus, it has no impact on the corresponding gravity theory.

\subsubsection{3.3.2 Matter dominated era}

In matter dominated era, $H=\frac{2}{3t}$, $\dot H=\frac{T}{4}$, the Ricci dark energy density is
\begin{equation}
\rho_D=-\frac{\alpha T}{4}.
\end{equation}
In order to solve \ref{gt3}, the value of $\omega_D$ needs to be known, this can be done by inserting expression of $\rho_D$ into \ref{conteq} first, which leads to
\begin{equation}
(4H\dot H + \ddot H)+3H(1+\omega_D)(2H^2+\dot H)=0.\label{conteq2}
\end{equation}
Then $\omega_D$ can be get by inserting $H=\frac{2}{3t}$ into \ref{conteq2}, the result is
\begin{equation}
\omega_D=0,
\end{equation}
which tells that the dark energy behaves the same as that of matter during the matter dominated era. In fact, since $\rho_D=3\alpha(2H^2+\dot H)\propto H^2$ in this special time duration, the same result can be get through similar argument as that in \cite{shu}. Now, one know that \ref{gt3} turns into
\begin{equation}
T^2g_{TT}+\frac{T}{2}g_T+\frac{\alpha T}{4}=0,
\end{equation}
whose solution is
\begin{equation}
g(T)=2C_1\sqrt {T}-\frac{1}{2}\alpha T+C_2,
\end{equation}
then, the form of $f(T)$ is
\begin{equation}
f(T)=(1-\frac{1}{2}\alpha)T+2C_1\sqrt {T}+C_2.\label{rdematter}
\end{equation}
By using boundary conditions to determine the integral constants, one then gets
\begin{equation}
f(T)=(1-\frac{\alpha}{2})T+\alpha\sqrt {T_0T}-\frac{\alpha}{2}T_0.\label{rdematterfinal}
\end{equation}

\subsubsection{3.3.3 Dark energy dominated era}

In the dark energy dominated era, a Hubble parameter with future singularity should be assumed due to the Phantom character of RDE, in this paper, the assumption is taken as that in \cite{bamba} with
\begin{equation}
H=h(t_s-t)^{-1},\label{rdeh}
\mathrm{}\end{equation}
with $h$ a positive constant. Then,
\begin{equation}
\dot H =-\frac{T}{6h},
\end{equation}
\begin{equation}
\ddot H=2h(-\frac{T}{6h^2})^{\frac{3}{2}},
\end{equation}
\begin{equation}
\rho_D=(\alpha+\frac{\alpha}{2h})T.
\end{equation}
A subtle point is that one should not use \ref{omegaga} to give the value of $\omega_D$ here. \ref{omegaga} is in fact a qualitative investigation of the state index when lack of the full knowledge of the scale factor. However, when the form of Hubble parameter $H$ is given in detail as \ref{rdeh}, the state index $\omega$ should also be given quantitatively. This can be done by inserting $\dot H$ and $\ddot H$ into \ref{conteq2}, then, $\omega_D$'s value is
\begin{equation}
\omega_D=-1-\frac{2}{3h},
\end{equation}
which is negative definitely, and thus is consistent with the assumption of Hubble parameter \ref{rdeh} with a future singularity. Now, \ref{gt3}
turns into
\begin{equation}
T^2g_{TT}+\frac{T}{2}+(\alpha+\frac{\alpha}{2h})T=0,
\end{equation}
whose solution is
\begin{equation}
g(T)=2C_1\sqrt T-(2\alpha+\frac{\alpha}{h})T+C_2.
\end{equation}
The form of $f(T)$ then is
\begin{equation}
f(T)=(1-2\alpha-\frac{\alpha}{h})T+2C_1\sqrt T+C_2.\label{rdede}
\end{equation}
Using boundary conditions to determine the integral constants, then, \ref{rdede} becomes to
\begin{equation}
f(T)=(1-2\alpha-\frac{\alpha}{h})T+(4\alpha+\frac{2\alpha}{h})\sqrt {T_0T}+(2\alpha+\frac{\alpha}{h})T_0.
\end{equation}

\section{4. Summary and discussion}

In the present paper, we have investigated the reconstruction of $f(T)$ theories from three different holographic dark energy models in various time durations. For the HDE model, the dark energy dominated era is chosen for reconstruction, and the radiation dominated era is chosen when the involved model changes into NADE. For the RDE model, however, radiation, matter and dark energy dominated time durations are all investigated.

The choice of particular time duration for $f(T)$ theory reconstruction is limited by the knowledge of the evolution of scale factor $a(t)$. This can be seen from the definition of dark energy density in HDE and NADE model, respectively. Let's take the reconstruction of $f(T)$ theory according to HDE model in a matter dominated era for an example. In this situation, in order to solve \ref{gt3} to get the resulting $f(T)$ theory, the evolution of $a(t)$ from matter to dark energy dominated era should be known quantitatively to define the dark energy density, however, a detail quantitative knowledge of this is not available at moment, which limit the reconstruction in this time duration. Situation is similar in NADE model. On contrary, in the RDE model, the definition of dark energy density doesn't depend on a global quantity, thus, the reconstruction can be done in any time duration.

It should be noted that, due to terms like $\ln(T)$ and $\sqrt T$ with torsion scalar $T$ negative definitely, the original forms of $f(T)$ (see \ref{solution11}, \ref{adeft}, \ref{rder}, \ref{rdematter} and \ref{rdede}),  all seem unacceptable because of breaking the hermiticity of Lagrangian density, however, by using boundary conditions similar to that in \cite{boundarycondition}, such problem is avoided. Thus, these boundary conditions do well in the sense that they give out result which are indeed physical expected. However, these boundary conditions seems not adequate enough to give more information of the corresponding $f(T)$ theory. An example can be discovered in the reconstruction of $f(T)$ theory from RDE model in matter dominated duration. Since $\omega_D=0$ there, the dark energy behaves the same as that of matter, the corresponding Friedmann equation should be just a rescaling of the original one, and correspondingly, $f(T)$ theory should also be a rescaling of the original TEGR, that is $h(\alpha)T$, $h(\alpha)$ here is a parameter which is a function of $\alpha$ in RDE. However, such expected information cannot be seen in \ref{rdematterfinal}. Thus, for further study, to get a more precise reconstruction of $f(T)$ theory from holographic dark energy models, some improved boundary conditions should be considered.

\acknowledgments

The work is partly supported by National Natural Science Foundation of China (No. 11275017 and No. 11173028).


\end{document}